\documentclass[12pt,a4paper]{iopart}

\usepackage{iopams}
\usepackage{graphicx,subfigure}

\newcommand{\msc}[1]{\vspace{10pt}
     \begin{indented}
     \item[]\rm Mathematics Subject Classification: #1\par
     \end{indented}}

\newcommand\binom[2]{\left(\begin{array}{c} #1 \\ #2 \end{array}
\right)}
\newcommand{\sgn}{\mathop{\mathrm{sgn}}\nolimits}

\begin{document}
%\noindent {\em J. Phys. A: Math. Gen.} {\bf 36} (2003) L53--L59

\letter{Nodal domain distributions for quantum maps }

\author{J P Keating\dag,\ F Mezzadri\dag\  and A G Monastra\ddag}

\address{\dag\ School of Mathematics, University of Bristol, University
Walk, Bristol, BS8 1TW, UK}

\address{\ddag\ Department of Physics of Complex Systems,
 The Weizmann Institute of Science, 76100 Rehovot, Israel}
\eads{\mailto{j.p.keating@bristol.ac.uk},
\mailto{f.mezzadri@bristol.ac.uk},
\mailto{alejandro.monastra@weizmann.ac.il}}

%\mydate{11 November 2002}

\begin{abstract}
The statistics of the nodal lines and nodal domains of the
eigenfunctions of quantum billiards have recently been observed to
be fingerprints of the chaoticity of the underlying classical
motion by Blum {\it et al} (2002 {\it Phys. Rev. Lett.} {\bf 88}
114101) and by Bogomolny and Schmit (2002 {\it Phys. Rev. Lett.}
{\bf 88} 114102). These statistics were shown to be computable
from the random wave model of the eigenfunctions. We here study
the analogous problem for chaotic maps whose phase space is the
two-torus. We show that the distributions of the numbers of nodal
points and nodal domains of the eigenvectors of the corresponding
quantum maps can be computed straightforwardly and exactly using
random matrix theory. We compare the predictions with the results
of numerical computations involving quantum perturbed cat maps.
\end{abstract}
\msc{81Q50, 15A52}

\vspace{.5cm} \noindent In a recent article Blum \etal (2002)
observed that the number-distributions of the nodal domains of
quantum wavefunctions of billiards whose classical dynamics is
integrable are different from those for chaotic billiards and
argued that the latter are universal. Thus, the
number-distribution of nodal domains appears to be a new criterion
for quantum chaos that complements the usual ones based on
spectral fluctuations. Blum \etal computed these distributions for
some integrable (and separable) systems, but no analytic formula
exists for the number of nodal domains of a chaotic billiard.
Berry (1977) has conjectured that the wavefunctions of quantum
systems with a chaotic classical limit behave like Gaussian random
functions. Supported by numerical evidence, Blum \etal found that
the limiting distribution of the number of nodal domains can be
reproduced assuming Berry's conjecture. Bogomolny and Schmit
(2002) developed a percolation model for nodal domains of Gaussian
random functions and showed that their number is Gaussian
distributed. They computed the mean and variance of this
distribution, which are both proportional to the mean spectral
counting function. Their results agree with the numerical
computations reported by Blum {\it et al} for chaotic billiards.
The influence of a boundary on the nodal lines of Gaussian random
functions has been investigated by Berry (2002), Gnutzmann \etal
(2002), and Berry and Ishio (2002).  This is expected to model the
nodal properties of billiard wavefunctions near boundaries.

We here consider the analogous problem for one-dimensional
time-reversal-symmetric systems with discrete time evolution and
whose phase space is the two-dimensional torus $\mathbb{T}^2$. The
classical dynamics of such systems corresponds to the action of
symplectic maps on $\mathbb{T}^2$, and their quantum mechanics to
that of unitary matrices $U_N$ (called propagators or quantum
maps) on a Hilbert space of dimension $N = 1/h$, where $h$ is
Planck's constant. Modelling the eigenvectors of $U_N$ by those of
random unitary symmetric matrices (such matrices constitute the
circular orthogonal ensemble, COE, of random matrix theory), we
compute the number-distributions of nodal domains and nodal points
(the analogues of nodal lines in billiards) exactly. It is shown
that these become Gaussian as $N \rightarrow \infty$ and that the
mean and variance are proportional to $N$ (precisely as in the
billiard case). We compare our results with numerical computations
involving the eigenvectors of perturbations of quantum cat maps
whose classical dynamics are hyperbolic and whose spectral
statistics are known to be accurately predicted by random matrix
theory (Basilio de Matos and Ozorio de Almeida 1995, Keating and
Mezzadri 2000).

Consider the Helmholtz equation with Dirichlet boundary conditions
\begin{equation}
\label{bileq} -\triangle
\Psi(\mathbf{r}) = E\, \Psi(\mathbf{r}),
\quad \mathbf{r} \in \Omega,
\end{equation}
where $\Omega$ is a connected compact domain in a two-dimensional
Riemann manifold. The {\it nodal lines} are the zero sets of real
solutions of equation~\eref{bileq}; the {\it nodal domains} are
connected domains in $\Omega$ where $\Psi(\mathbf{r})$ has
constant sign.  Now, let $\{\Psi_n(\mathbf{r})\}_{n=1}^\infty$ be
a set of eigenfunctions of the laplacian on $\Omega$ ordered by
the magnitude of the corresponding eigenvalue $E_n$, and let
$\nu_n$ be the number of nodal domains of the $n$-th
eigenfunction. Courant (1923) proved that $ \nu_n \le n$.  Let
$I_g(E)=[E, E +gE]$, for $g > 0$. Blum \etal (2002) introduced the
distribution
\begin{equation}
P_{{\rm b}}(x,I_g(E))= \frac{1}{N_I} \sum_{E_n \in I_g(E)}
\delta\left(x - \frac{\nu_n}{n}\right),
\end{equation}
where $N_I$ is the number of energy levels in $I_g(E)$.  The
limiting distribution of nodal domains is defined by
\begin{equation}
\label{dndb} P_{{\rm b}}(x) = \lim_{E \rightarrow \infty} P_{{\rm
b}}(x,I_g(E)).
\end{equation}

We now introduce a density that is the analogue of~\eref{dndb} for
quantum maps. The periodicity of the two-torus constrains the
wavefunction to be an infinite sum of delta-functions supported at
rational points of the form $j/N$, with $j$ integer, in both the
position and momentum basis (Hannay and Berry 1980), i.e.
\numparts
\begin{eqnarray}
\psi(q) & = \sum_{m \in \mathbb{Z}} \sum_{j=1}^{N} c_j \, \delta
\left(q - \frac{j}{N} + m\right), \\
\hat{\psi}(p) & = \sum_{m \in \mathbb{Z}} \sum_{j=1}^{N} \hat{c}_j
\, \delta \left(p - \frac{j}{N} + m\right),
\end{eqnarray}
\endnumparts
where $N = 1/h$ and
\begin{equation}
\hat{\psi}(p) = \frac{1}{\sqrt{2 \pi \hbar}}
\int_{-\infty}^{\infty} \psi(q)\, \rme^{-\frac{\rmi q p}{\hbar}}\;
\rmd q.
\end{equation}
Moreover, since $\psi(q)$ and $\hat{\psi}(p)$ are periodic,
$c_j=c_{j + N}$ and $\hat{c}_j = \hat{c}_{j + N}$. Therefore, a
quantum state is completely determined by $N$ complex numbers,
which implies that the Hilbert space is isomorphic to
$\mathbb{C}^N$. The coefficient $c_j$ can thus be interpreted as
the value of $\psi(q)$ at $q=j/N$ (the Heisenberg uncertainty
principle is not violated, because the periodic sum of
delta-functions that defines $\hat{\psi}(p)$ extends to infinity).
Now, let $U_N$ be the matrix realization of a quantum map in the
basis $\{ | \, j \,\rangle \}_{j=1}^{N}$, where
\begin{equation}
\label{bq}
\langle \, q \, | \, j \, \rangle = \sum_{m \in
\mathbb{Z}} \, \delta \left( q - \frac{j}{N} + m\right).
\end{equation}
We shall consider only systems whose dynamics is invariant under
time reversal, so that $U_N$ is a symmetric unitary matrix and,
without loss of generality, the eigenvectors can be taken to be
real.

Because of the topology of the phase space, an eigenvector of
$U_N$ is equivalent to a sequence of $N$ real numbers with
periodic boundary conditions, i.e. $c_1 = c_{N + 1}$. A nodal
point is then identified whenever two consecutive coefficients
$c_j$ have opposite sign. The total number of nodal points in a
given eigenvector is
\begin{equation}
\nu = \frac{1}{2} \sum_{j=1}^{N} [ 1 - \sgn (c_j) \sgn (c_{j+1})
],
\end{equation}
where
\begin{equation}
\sgn(x) = \cases{ 1 & if $x>0$, \\
                   0 & if $x=0$, \\
                   -1 & if $x< 0$.}
\end{equation}
Similarly, a nodal domain is a set of consecutive integers
$\{j+1,j+2,\ldots,j+k\}$ such that the corresponding coefficients
$c_j$ lie between two nodal points and thus have constant sign. As
a consequence of the periodicity of the coefficients $c_j$, there
can be only an even number of nodal points, equal to the number of
nodal domains; the only exception is when there are no nodal
points and only one nodal domain. It follows from the results to
be presented later that as $N \rightarrow \infty$ the probability
that all the $c_j\,$s have the same sign is negligible, and so we
shall denote by $\nu$ both the number of nodal points and the
number of nodal domains. Finally, the limiting distribution is
defined by
\begin{equation}
\label{mdis}
 P_{{\rm m}}(x) = \lim_{N \rightarrow \infty}
\frac{1}{N}\sum_{n=1}^{N} \, \delta\left(x -
\frac{\nu_n}{N}\right),
\end{equation}
where, as for billiards, $\nu_n$ is the number of nodal domains
(points) of the $n$-th eigenvector. Identical definitions can
obviously be formulated in the momentum representation.

When the classical limit of $U_N$ is a chaotic map, the
eigenstatistics of $U_N$ are expected to be the same as those of
matrices in the COE (Bohigas {\it et al} 1984).  The COE
probability measure is invariant under the mapping
\begin{equation}
\label{trans}
 U \mapsto OUO^T,
\end{equation}
where $U$ is a unitary symmetric matrix and $O$ is an arbitrary
orthogonal matrix. Hence, each eigenvector of $U$ is mapped by an
orthogonal transformation into an eigenvector of a new matrix that
by~\eref{trans} has the same weight in the ensemble as $U$ and the
same spectrum. As a consequence (see, e.g. Haake 2000), the
eigenvectors of matrices in the COE are uniformly distributed on
the unit sphere in $\mathbb{R}^N$ and the joint probability
density of their components is
\begin{equation}
\label{jpdc} P_{{\rm
COE}}(c_1,c_2,\ldots,c_N)=\frac{1}{2\pi^{N/2}} \,
\Gamma\left(\frac{N}{2}\right) \delta\left(1 -\sum_{j=1}^{N}
c_j^2\right).
\end{equation}
The above distribution is independent of the signs of the
$c_j\,$s, therefore they can be either positive or negative with
equal probability and there are no correlations among the signs of
different coefficients. This simple observation allows us to
compute analytically all the relevant quantities in a very
straightforward way.

The signs of the $c_j\,$s behave like a sequence of $N$
independent random variables $s_j$ that assume the values
$\{1,-1\}$ with equal probability $\frac12$; in other words, they
are equivalent to an array of non-interacting particles  with spin
$\frac12$ and periodic boundary conditions. Thus, the probability
of a configuration with $N_+$ spins up and $N_{-}=N-N_+$ spins
down is given by the binomial distribution
\begin{equation}
P(N_+,N_{-})=\frac{1}{2^N} \binom{N}{N_+}.
\end{equation}
The computation of the density~\eref{mdis} requires a simple
combinatorial argument.  In a periodic chain of $N$ spins there
are $N$ possible positions where a nodal point can be located.
Hence, the number of configurations with $\nu$ nodal points is
zero when $\nu$ is odd and twice the number of ways of choosing
$\nu$ objects among $N$, irrespective of their ordering, for even
$\nu$, i.e.
\begin{equation}
[1 + (-1)^\nu] \binom{N}{\nu}.
\end{equation}
The factor of two in front of the binomial coefficient is due to
the fact that by simultaneously changing the sign of all the spins
in the chain we obtain a new configuration with the nodal points
in the same positions. Finally, the distribution of the number of
nodal points and nodal domains is given by
\begin{equation}
\label{ndc}
P_{{\rm m}}(\nu, N) = \frac{1 +
(-1)^\nu}{2^{N}}\binom{N}{\nu}.
\end{equation}
The mean $\langle \nu \rangle$ and variance $\sigma^2=\langle
\nu^2 \rangle - \langle \nu \rangle^2$ can be easily computed:
\begin{equation}
\label{m&v}
 \langle \nu \rangle  = \frac{N}{2} \quad {\rm and}
\quad \sigma^2 = \frac{N}{4}.
\end{equation}
Equations~\eref{ndc} and~\eref{m&v} correspond to the results that
Bogomolny and Schmit (2002) obtained for the percolation model of
random wave functions in two-dimensional systems. By letting $N
\rightarrow \infty$ and scaling $x = \nu/N$, the discrete
distribution~\eref{ndc} tends to a continuous Gaussian probability
density with mean $\frac12$ and variance $\sigma^2 = 1/4N$, i.e.
\begin{equation}
\label{limitgaus}
 P_{{\rm m}}(x,N) \sim
 \sqrt{\frac{2N}{\pi}}\, \exp\left[-2N \left(x -
 1/2\right)^2\right], \quad N \rightarrow \infty.
\end{equation}
This is the main result of this note.

\begin{figure}
\centering
\subfigure[]{\label{fig1a}
\includegraphics[width=3.0in]{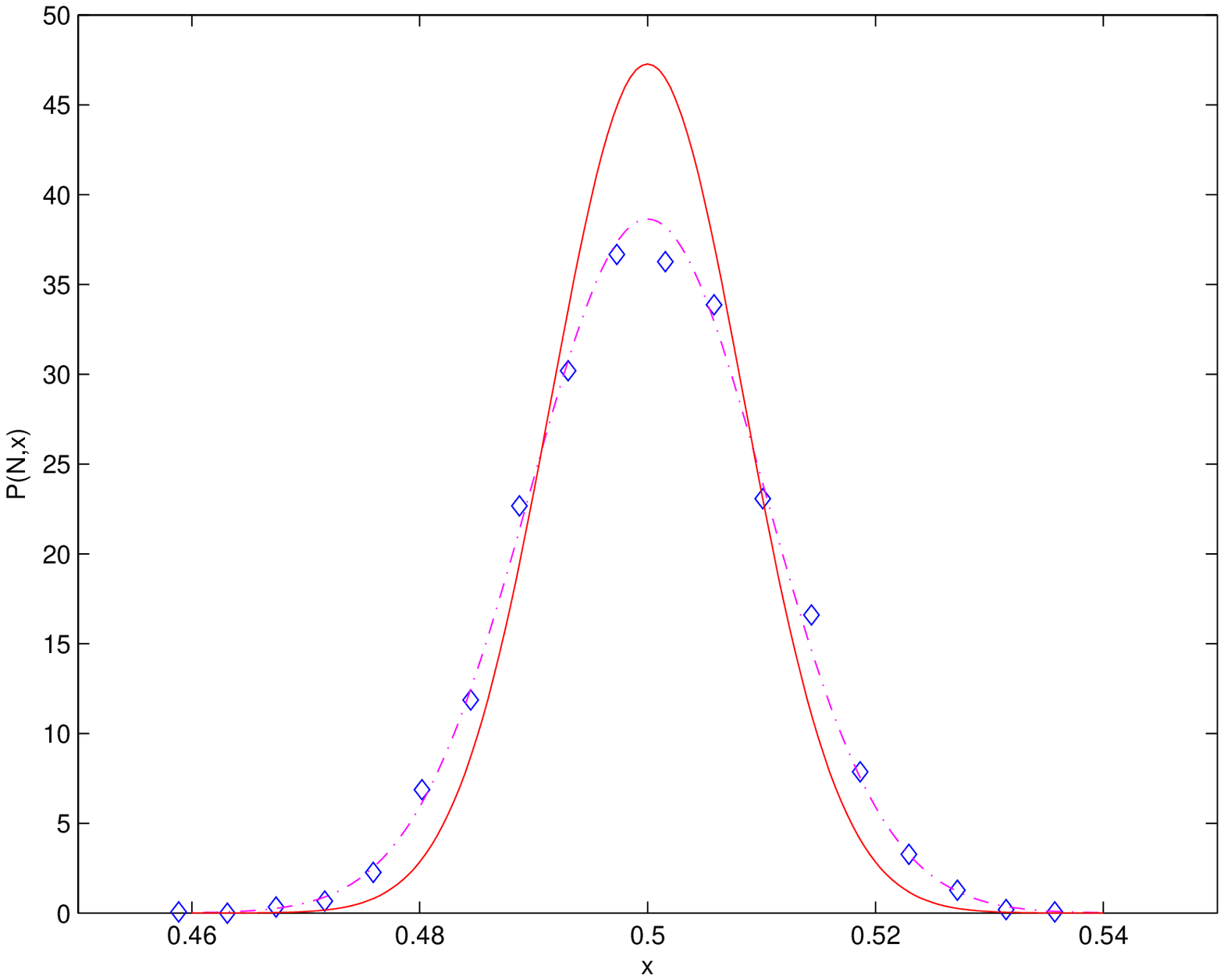}}
 \subfigure[]{\label{fig1b}
  \includegraphics[width=3.0in]{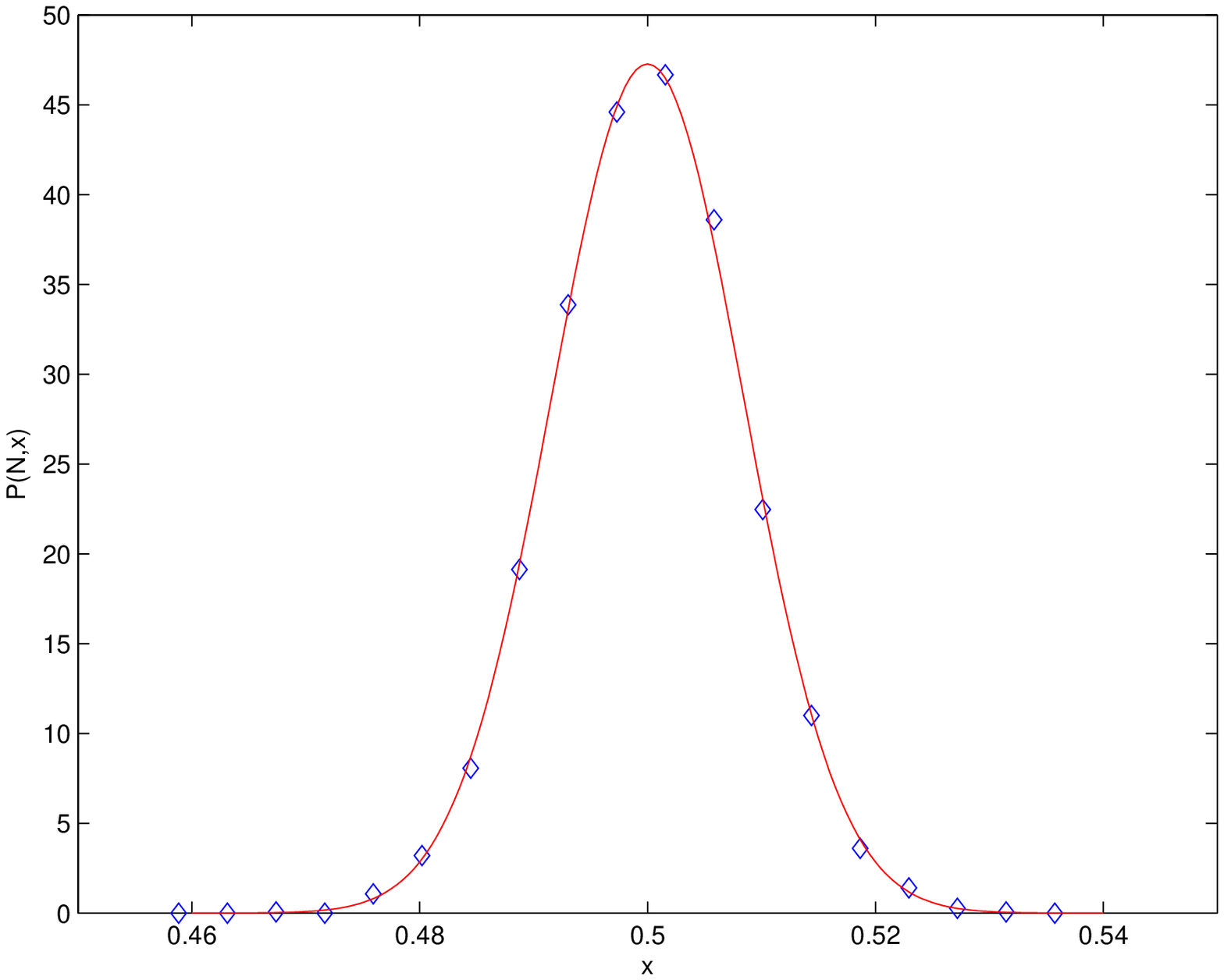}}
      \caption{Nodal domain distributions (\opendiamond) of the
      unperturbed quantum cat map (a) and of its
      perturbation~\eref{mephi} with $k=0.30$ (b) compared with
      the Gaussian~\eref{limitgaus}
      (\full). The dimension of the Hilbert space is $N=3511$.}
      \label{fig1}
\end{figure}
In order to compare the distribution~\eref{limitgaus} with
numerical computations, we consider perturbations of the following
hyperbolic (cat) map:
\begin{equation}
\label{cm}
A: \binom{q}{p} \mapsto \left( \begin{array}{cc} 2 & 1 \\ 3 & 2
\end{array}\right) \binom{q}{p} \quad \bmod 1.
\end{equation}
Because of the number-theoretical properties of $A$, the spectrum
of the propagator $U_N(A)$ is non-generic (Keating 1991, Kurlberg
and Rudnick 2000) in that it does not obey the random matrix
theory conjecture. However, if a small nonlinear perturbation is
introduced, the composite map is still hyperbolic but loses its
arithmetical nature. As a consequence, the spectrum of the new
quantum map has random matrix correlations. Hence, we
perturb~\eref{cm} with the following shear in the momentum
\begin{equation}
\label{shear}
\brho: \binom{q}{p} \mapsto \binom{q}{p +
\frac{k}{4\pi}\cos (2\pi q)}
\end{equation}
and study the propagator $U_N(\bphi)$ of the map
\begin{equation}
\label{pert} \bphi = \brho \circ A \circ \brho.
\end{equation}
The matrix elements of this propagator in the basis~\eref{bq} are
\begin{equation}
\label{mephi} \fl
U_N(\bphi)_{l \,m} = \frac{1}{\sqrt{\rmi N}}\,
\exp\left\{ \frac{2\pi \rmi}{N} \left[l^2 - lm + m^2 +
\frac{N^2k}{8\pi^2}\left(\sin(2\pi l / N) + \sin(2\pi
m/N)\right)\right]\right\}
\end{equation}
(Basilio de Matos and Ozorio de Almeida 1995). It can be shown
that the only symmetry of this quantum map is time reversal
(Keating and Mezzadri 2000). Furthermore, if $k < k_{{\rm max}} =
0.32\ldots $, then the map~\eref{pert} is uniformly hyperbolic and
the spectral statistics of the propagator~\eref{mephi} are
consistent with random matrix theory  (Basilio de Matos and Ozorio
de Almeida 1995). Figure~\ref{fig1b} shows the nodal domain
distribution of the eigenvectors of the quantum map~\eref{mephi}
for a particular choice of $k$ and $N$, together with the
density~\eref{limitgaus}. The nodal domain distribution of the
unperturbed quantum map, figure~\ref{fig1a}, also appears to be
Gaussian, but its variance cannot be predicted by random matrix
theory.

\begin{figure}
\centering
\subfigure[]{
\includegraphics[width=3.0in]{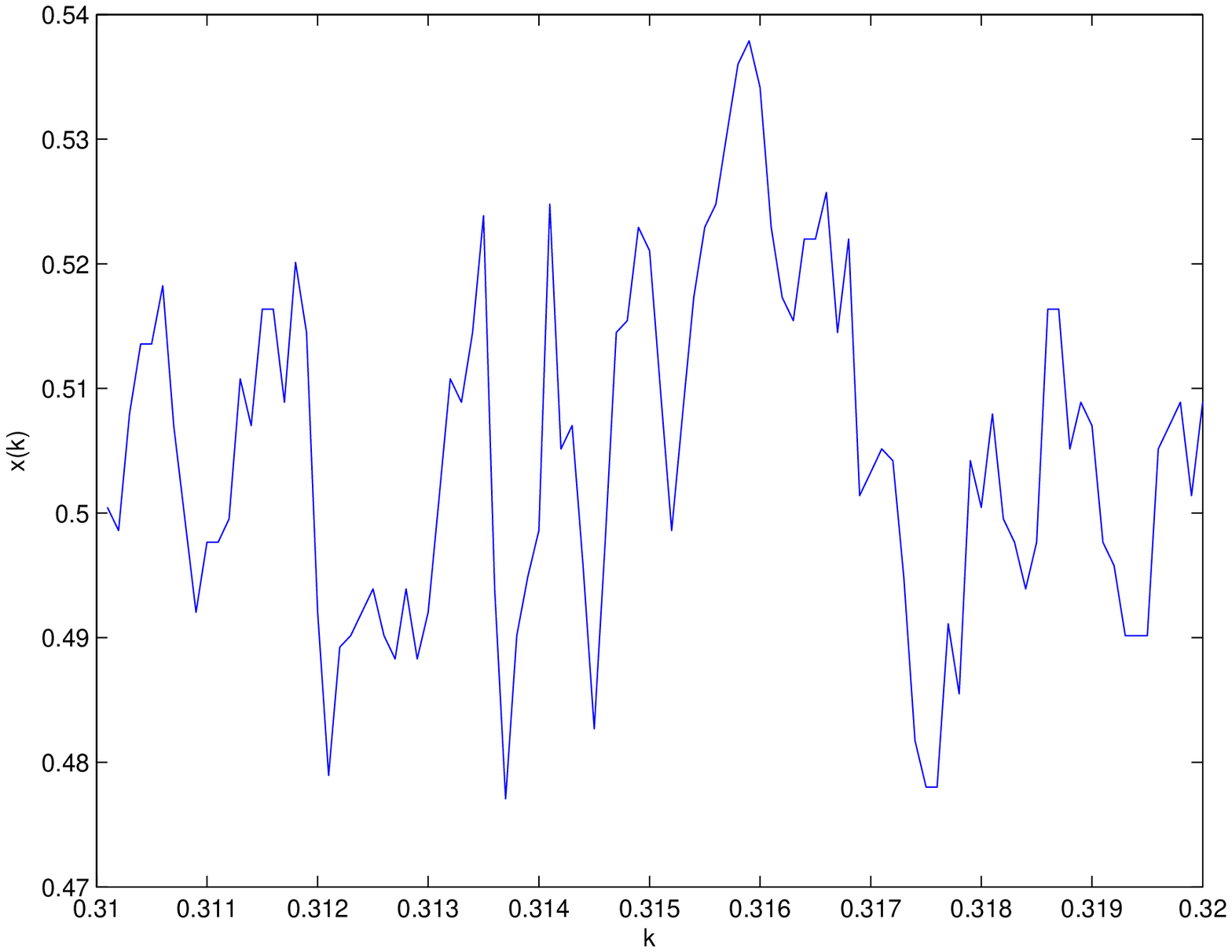}}
 \subfigure[]%
  {\includegraphics[width=3.0in]{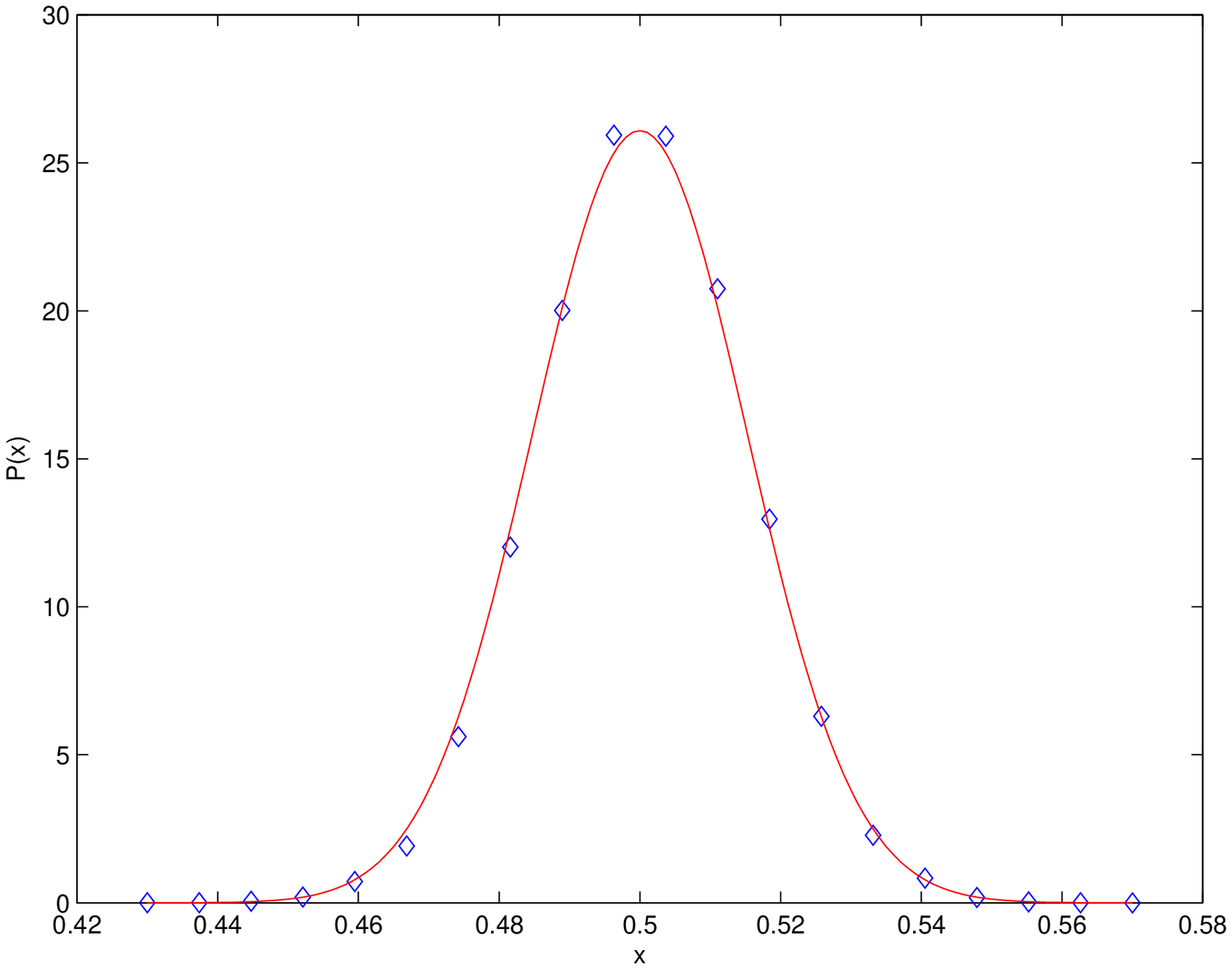}}
      \caption{(a) Scaled number of nodal domains $x(k)=\nu(k)/N$ of an
        eigenvector of the matrix~\eref{mephi}, with $N=1069$,
         as a function of the perturbation parameter $k$;
        (b) value distribution of $x(k)$ averaged over all
        eigenvectors (\opendiamond) compared with the
         Gaussian~\eref{limitgaus} (\full).}
      \label{cod}
\end{figure}

As the perturbation parameter $k$ varies, the nodal points in a
given eigenvector of the matrix~\eref{mephi} change their
positions. A natural question then arises: what is the minimum
number of parameters needed to create or coalesce nodal points and
alter the number of nodal domains? In other words, what is the
codimension of the nodal points? Since the spins in a chain are
uncorrelated, the functions $s_j(k)$ will be independent,
therefore nodal points move randomly without repelling or
attracting each other. Thus, the codimension of nodal points is
one and a single parameter is enough to create or annihilate nodal
domains with equal probability.  This behaviour is illustrated in
figure~\ref{cod}; the scaled number of nodal domains $x(k)=
\nu(k)/N$ of an eigenvector oscillates around $\frac12$, and since
the $s_j(k)$ are independent, the value distribution of $x(k)$ is
given by the Gaussian~\eref{limitgaus}.

Finally, it is worth remarking that this problem is equivalent to
a one-dimensional Ising model of {\it non-interacting} spins in a
magnetic field $B$ with periodic boundary conditions, whose
Hamiltonian and partition function are
\begin{equation}
\label{Isingmod}
 H=-B\sum_{j=1}^{N}s_j, \quad s_j = \pm 1, \quad s_1=s_{N+1}
\end{equation}
and
\begin{equation}
Z(\beta,B)=\sum_{\{s_1\}}\sum_{\{s_2\}} \ldots \sum_{\{s_N\}}
\exp\left(- \beta H\right) = 2^N\cosh(\beta B)^N
\end{equation}
respectively. All the relevant thermodynamical quantities should
be computed at $\beta= B = 0$. This plays the role in this case of
the analogy between the nodal statistics of billiard wavefunctions
and the Potts model suggested by Bogomolny and Schmit (2002).\\

\ack
FM was supported by a Royal Society Dorothy Hodgkin Research
Fellowship.  AGM is grateful to the EU network ``Mathematical
Aspects of Quantum Chaos''  for financial support during his visit
to the University of Bristol, where this work was carried out.

\section*{References}
\begin{harvard}
\item[]Basilio de Matos M and Ozorio de Almeida A M 1995
Quantization of Anosov maps {\it Ann. Phys. NY} {\bf 237} 46--65
\item[]Berry M V 1977 Regular and irregular semiclassical wave
functions {\it J. Phys. A: Math. Gen.} {\bf 10} 2083--91
\item[]\dash 2002 Statistics of nodal lines and points in chaotic
quantum billiards: perimeter corrections, fluctuations, curvature
{\it J. Phys. A: Math. Gen.} {\bf 35} 3025--38 \item[]Berry M V
and Ishio H 2002 Nodal densities of Gaussian random waves
satisfying mixed boundary conditions {\it J. Phys. A: Math. Gen.}
{\bf 35} 5961--72 \item[]Blum G, Gnutzmann S and Smilansky U 2002
Nodal Domains Statistics: A Criterion for Quantum Chaos {\it Phys.
Rev. Lett.} {\bf 88} 114101 \item[]Bogomolny E and Schmit C 2002
Percolation Model for Nodal Domains of Chaotic Wave functions {\it
Phys. Rev. Lett.} {\bf 88} 114102 \item[]Bohigas O, Giannoni M J
and Schmit C 1984 Characterization of Chaotic Quantum Spectra and
Universality of Level Fluctuation Laws {\it Phys. Rev. Lett.} {\bf
52} 1--4 \item[]Courant R 1923 Ein allgemeiner Satz zur Theorie
der Eigenfunktione selbstadjungierter Differentialausdr\"ucke {\it
Nach. Ges. Wiss. G\"ottingen Math.-Phys. Kl. (G\"ottingen, 13 July
1923)} pp~81--4 \item[]Gnutzmann S, Monastra A G and Smilansky U
2002 Avoided intersections of nodal lines {\it Preprint}
nlin.CD/0212006 \item[]Haake F 2000 {\it Quantum Signature of
Chaos} (Berlin, Heidelberg and New York: Springer-Verlag) p 60
\item[]Hannay J H and Berry M V 1980 Quantization of linear maps
on the torus --- Fresnel diffraction by a periodic grating {\it
Physica} D{\bf 1} 267--90 \item[] Keating J P 1991 Asymptotic
properties of the periodic orbits of the cat maps {\it
Nonlinearity} {\bf 4} 277--307 \item[]\dash 1991 The cat maps:
quantum mechanics and classical motion {\it Nonlinearity} {\bf 4}
309--41 \item[] Keating J P and Mezzadri F 2000 Pseudo-symmetries
of Anosov maps and spectral statistics {\it Nonlinearity} {\bf 13}
747--75 \item[] Kulberg P and Rudnick Z 2000 Hecke theory and
equidistribution for the quantization of linear maps of the torus
{\it Duke Math. J.} {\bf 103} 47--78
\end{harvard}
\end{document}